\documentstyle[12pt,epsfig,amsfonts]{article}
\makeatletter

\@addtoreset{equation}{section} \makeatother
\newcommand{\nn}{\nonumber}
\newcommand{\be}{\begin{equation}}
\newcommand{\ee}{\end{equation}}
\newcommand{\ba}{\begin{eqnarray}}
\newcommand{\ea}{\end{eqnarray}}
\newcommand{\dle}[1]{\label{#1}}

\newcommand{\dr}[1]{\ref{#1}}
\newcommand{\dc}[1]{\cite{#1}}

\newcommand{\gsim}{\raise.3ex\hbox{$>$\kern-.75em\lower1ex\hbox{$\sim$}}}
\newcommand{\lsim}{\raise.3ex\hbox{$<$\kern-.75em\lower1ex\hbox{$\sim$}}}

\newcommand{\si}{{\sigma}}

\newcommand{\al}{\alpha}

\newcommand{\ga}{\gamma}

\newcommand{\La}{\Lambda}
\newcommand{\la}{\lambda}
\newcommand{\Om}{\Omega}

\newcommand{\vph}{\varphi}

\newcommand{\dd}{\partial}

\newcommand{\dR}{\dot{R}}
\newcommand{\dRR}{\dot{R}^2}
\newcommand{\vpk}{\varphi_k}
\newcommand{\vpkt}{\varphi_{k,\tau \tau}}
\newcommand{\tE}{\tilde{E}}
\newcommand{\pk}{\phi_k}
\newcommand{\cH}{{\cal{H}}}
\newcommand{\cM}{{\cal{M}}}

\begin{document}

\renewcommand{\thefootnote}{\fnsymbol{footnote}}
\begin{flushright}
ORSAY-LPT-02-62 hep-th/0206147
\end{flushright}
\begin{center}

{\large\bf Perturbations on a moving D3-brane and mirage
cosmology} \vskip 0.1cm \vskip 1.2cm Timon
Boehm$^{a}$\footnote{{\tt Timon.Boehm@physics.unige.ch}} and
D.A.Steer$^{b}$\footnote{{\tt steer@th.u-psud.fr}}
\\
\vskip 10pt {\it a})
D\'epartement de Physique Th\'eorique, Universit\'e de Gen\`eve,\\
24 quai E. Ansermet, CH--1211 Geneva 4, Switzerland. \vskip 3pt
{\it b}) Laboratoire de Physique Th\'eorique,
B\^at 210, Universit\'e Paris XI, \\
Orsay Cedex, France and\\
F\'ed\'eration de Recherche APC, Universit\'e Paris VII, France.
\\
\end{center}
\vskip 1.2cm

\renewcommand{\thefootnote}{\arabic{footnote}}
\setcounter{footnote}{0} \typeout{--- Main Text Start ---}
\begin{abstract}

We study the evolution of perturbations on a moving probe D3-brane
coupled to a 4-form field in an AdS$_5$-Schwarzschild bulk.  The
unperturbed dynamics are parameterized by a conserved energy $E$
and lead to Friedmann-Robertson-Walker `mirage' cosmology on the
brane with scale factor $a(\tau)$. The fluctuations about the
unperturbed worldsheet are then described by a scalar field
$\phi(\tau,\vec{x})$. We derive an equation of motion for $\phi$,
and find that in certain regimes of $a$ the effective mass squared
is negative. On an expanding BPS brane with $E=0$ superhorizon
modes grow as $a^4$ whilst subhorizon modes are stable. When the
brane contracts, all modes grow. We also briefly discuss the case
when $E>0$, BPS anti-branes as well as non-BPS branes. Finally,
the perturbed brane embedding gives rise to scalar perturbations
in the FRW universe.  We show that $\phi$ is proportional to the
gauge invariant Bardeen potentials on the brane.
\end{abstract}


\section{Introduction}

The idea that our universe may be a 3-brane embedded in a higher
dimensional space-time is strongly motivated by string- and
M-theory, and it has recently received a great deal of attention.
Much work has focused on the case in which the universe 3-brane is
of co-dimension 1
\dc{Rubakov:1983bb,Randall:1999vf,Randall:1999ee} and the
resulting cosmology (see e.g.\
\dc{Binetruy:1999ut,Cline:1999ts,Csaki:1999jh}) and cosmological
perturbation theory (e.g.\
\dc{vandeBruck:2000ju,Mukohyama:2001yp,Langlois:2000ia,Riazuelo:2002mi,
Bridgman:2001mc,Kodama:2000fa,Koyama:2001ct}) have been studied in
depth. When there is more than one extra dimension the Israel
junction conditions, which are central to the 5D studies, do not
apply and other approaches must be used
\dc{Burgess:2001fx,Kehagias:1999vr,Alexander:2001ks}. In the
`mirage' cosmology approach \dc{Kehagias:1999vr,Kiritsis:1999tj}
the bulk is taken to be a given supergravity solution, and our
universe is a {\it test} D3-brane which moves in this background
spacetime so that its back-reactions onto the bulk is neglected.
If the bulk metric has certain symmetry properties, the
unperturbed brane motion leads to FRW cosmology with scale factor
$a(\tau)$ on the brane \dc{Kehagias:1999vr,Steer:2002ng}. Our aim
in this paper is to study the evolution of perturbations on such a
moving brane. Given the probe nature of the brane, this question
has many similarities with the study of the dynamics and
perturbations of cosmic topological defects such as cosmic strings
\dc{Hindmarsh:1995re,Garriga:1991ts,Guven:1993ew,Battye:1998zk}.


Though we derive the perturbation equations in a more general
case, we consider in the end a bulk with
AdS$_5$-Schwarzschild$\times$S$_5$ geometry which is the near
horizon limit of the 10-dimensional black D3-brane solution. In
this limit (using the AdS-CFT correspondence)  black-hole
thermodynamics can be studied via the probe D3-brane dynamics
\dc{Cai:1999ad,Savonije:2001nd}. As discussed in section
\dr{subsec:setup}, we  make the assumption that the D3-brane has
no dynamics around the S$_5$ so that the bulk geometry is
effectively AdS$_5$-Schwarzschild. Due to the generalized Birkhoff
theorem \dc{Bowcock:2000cq}, this 5D geometry plays an important
r\^ole in work on co-dimension 1 brane cosmology. Hence links can
be made between the unperturbed probe brane FRW cosmology
discussed here and exact brane cosmology based on the junction
conditions \dc{Steer:2002ng}. Similarly the perturbation theory we
study here is just one limit of the full, self-interacting and
non-$Z_2$-symmetric brane perturbation theory which has been
studied elsewhere \dc{Riazuelo:2002mi}. Comments will be made in
the conclusions regarding generalizations of this work to the full
10D case.

Regarding the universe brane, the zeroth order (or background)
solution is taken to be an infinitely straight brane whose motion
is now constrained to be along the single extra dimension labelled
by coordinate $r$. The brane motion is parameterized by a
conserved positive energy $E$ \dc{Kehagias:1999vr}. In
AdS$_5$-Schwarzschild and to an observer on the brane, the motion
appears to be FRW expansion/contaction with a scale factor given
by $a \propto r$. Both the perturbed and unperturbed brane
dynamics will be obtained from the Dirac-Born-Infeld action for
type IIB superstring theory (see e.g.~\dc{Bachas:1998rg}),
\be
     S_{D3}=-T_3 \int d^{4}\si
                \sqrt{-\mbox{det}(\hat{\ga}_{ab}+2\pi \al'F_{ab}+\hat{B}_{ab})}
                -\rho_3 \int d^4\si \hat{C}_4.
\dle{eq:dbiaction}
\ee
Here $\sigma^{a}$ ($a=0,1,2,3$) are coordinates on the brane
worldsheet, $T_3$ is the brane tension, and in the second
Wess-Zumino term $\rho_3$ is the brane charge under a RR 4-form
field living in the bulk. We will write
\be
\rho_3 = q T_3
\ee
so that $q=(-)1$ for BPS (anti-)branes. In (\dr{eq:dbiaction})
$\hat{\ga}_{ab}$ is the induced  metric and $F_{ab}$  the field
strength tensor of the gauge fields on the brane. The quantities
$\hat{B}_{ab}$ and $\hat{C}_4$ are is the pull-backs of the
Neveu-Schwarz 2-form, and the Ramond-Ramond 4-form field in the
bulk. In the background we consider, the dilaton is a constant and
we set it to zero.  In general the brane will not move slowly, and
hence the square root in the DBI part of (\dr{eq:dbiaction}) may
not be expanded: we will consider the full non-linear action.
Finally, notice that since the 4D Riemann scalar  does not appear
in (\dr{eq:dbiaction}) (and it is not inherited from the
background in this probe brane approach) there is no brane
self-gravity. Hence the `mirage' cosmology we discuss here is
solely sourced by the brane motion, and it leads to effects which
are not present in 4-dimensional Einstein gravity. The lack of
brane self-gravity is a serious limitation. However, in certain
cases it may be included, for instance by compactifying the
background space-time as discussed in \dc{Brax:2002qw} (see also
\dc{Burgess:2001fx}). Generally this leads to bi-metric theories.
Even in that case, the mirage cosmology scale factor $a(\tau)$
which we discuss below plays an important r\^ole and hence we
believe it is of interest to study perturbations in this `probe
brane' approach.

Deviations from the infinitely straight moving brane give rise to
perturbations around the FRW solution.  Are these `wiggles'
stretched away by the expansion, or on the contrary do they grow
leading to instabilities?  To answer this question, we exploit the
similarity with uncharged cosmic topological defects and make use
of the work developed in that context by Garriga and Vilenkin
\dc{Garriga:1991ts}, Guven \dc{Guven:1993ew} and Battye and Carter
\dc{Battye:1998zk}.  The perturbation dynamics are studied through
a scalar field $\phi(\sigma)$ whose equation of motion is derived
from action (\dr{eq:dbiaction}).  We find that for an observer
comoving with the brane, $\phi$ has a tachyonic mass in certain
ranges of $r$ which depend on the conserved energy $E$
characterizing the unperturbed brane dynamics.  We discuss the
evolution of the modes $\phi_k$  for different $E$ and show that
in many cases the brane is unstable. In particular, both sub- and
super-horizon modes grow for a brane falling into the black hole.
It remains an open question to see if brane self-gravity,
neglected in this approach, can stabilize the system.

Finally, we also relate $\phi$ to the standard 4D  gauge invariant
scalar Bardeen potentials $\underline{\Phi}$ and
$\underline{\Psi}$ on the brane.  We find that $\underline{\Phi}
\propto \underline{\Psi} \propto \phi$ (no derivatives of $\phi$
enter into the Bardeen potentials).

The work presented here has some overlap with that of Carter et al
\dc{Carter:2001af} who also studied perturbations on moving
charged branes in the limit of negligible self-gravity.  Their
emphasis was on trying to mimic gravity on the brane, and in
addition they included matter on the brane.  Here we consider the
simplest case in which there is no matter on the brane: namely
$F_{ab}=0$ in (\dr{eq:dbiaction}). Our focus is on studying the
evolution of perturbations solely due to motion of the brane: we
expect the contribution of these perturbations to be important
also when matter is included. Moreover, we hope that this study
may more generally be of interest for the dynamics and
perturbations of moving D-branes in non-BPS backgrounds.

The outline of the paper is as follows. In section
\ref{sec:unperturbed} we link our 5-dimensional metric to the
10-dimensional black D3-brane solution and specify the unperturbed
embedding of the probe brane.  To determine its dynamics from the
action (\dr{eq:dbiaction}) the bulk 4-form  RR field must be
specified. We discuss the normalization of this field.  At the end
of the section we summarize the motion of the probe brane  by
means of an effective potential.  Comments are made regarding the
Friedmann equation for an observer on the brane. In section
\ref{sec:perturbed} we consider small deviations from the
background brane trajectory and investigate their evolution. The
equation of motion for $\phi$ is derived, and we solve it in
various regimes commenting on the resulting instabilities.  In
section \ref{sec:bardeen} we link  $\phi$ to the scalar Bardeen
potentials on the brane.  Finally, in section
\ref{sec:conclusions} we summarize our results.

\section{Unperturbed dynamics of the D3-brane}
\dle{sec:unperturbed}

In this section we discuss the background metric, briefly review
the unperturbed D3-brane dynamics, and comment on the cosmology as
seen by an observer on the brane.  The reader is referred to
\dc{Kehagias:1999vr,Papantonopoulos:2000yz} for a more detailed
analysis on which part of this section is based.

\subsection{Background metric and brane scale-factor} \label{subsec:setup}

For the reasons mentioned in the introduction, we focus mainly on
a AdS$_5$-S$\times$S$^5$ bulk spacetime. This is closely linked to
the 10D black 3-brane supergravity solution
\dc{Horowitz:1991cd,Aharony:1999ti,Kiritsis:1999tx} which
describes $N$ coincident D3-branes carrying RR charge $Q=N T_3$
and which is given by
\be
  ds_{10}^2=H_3^{-1/2}(-Fdt^2+d\vec{x}\cdot d\vec{x})
  +H_3^{1/2}\left(\frac{dr^2}{F}+r^2 d\Om_{5}^{2}\right)
  \dle{eq:sugrametric}
\ee
where the coordinates $(t,\vec{x})$ are parallel to the $N$
D3-branes, $d\Om_{5}^{2}$ is the line element on a 5-sphere and
\be
  H_3(r)=1+\frac{\ell^{4}}{r^{4}}, \qquad
  F=1-\frac{r_H^{4}}{r^{4}}.
  \dle{eq:harmonic}
\ee
The quantity  $\ell$ is the AdS$_5$ curvature radius and the
horizon  $r_H$ vanishes when the ADM mass equals $Q$. The link
between the metric parameters $\ell, r_H$ and the string
parameters $N, T_3$ is given e.g.~in \dc{Kiritsis:1999tx}.
The corresponding bulk RR field may also
be found in \dc{Kiritsis:1999tx}.

The near horizon limit of metric (\ref{eq:sugrametric}) is
AdS$_5$-S$\times$S$^5$ space time \dc{Aharony:1999ti}. Our
universe is taken to be a D3-brane moving in this background. We
make the following two assumptions.  First, the universe brane is
a probe so that its backreaction on the bulk geometry is
neglected. This may be justified if $N \gg 1$. Secondly, the probe
is assumed to have no dynamics around S$^5$ so that it is
constrained to move only along the radial direction $r$.  This is
a consistent solution of the {\it unperturbed} dynamics since the
brane has a conserved angular momentum about the S$^5$, and this
may be set to zero \dc{Kehagias:1999vr,Steer:2002ng}.  In section
\ref{sec:perturbed} we assume  that is also true for the perturbed
dynamics. Thus in the remainder of this paper we consider an
AdS$_5$-S bulk spacetime with metric
\ba
    ds_5^2 & = & -f(r)dt^2+ g(r)d\vec{x}\cdot d\vec{x}+h(r)dr^2
\dle{5D}\\
& \equiv & g_{\mu \nu} dx^{\mu} dx^{\nu}
\dle{eq:adsssmetric}
\ea
where
\be
    f(r) = \frac{r^2}{\ell^2} \left( 1 - \frac{r_H^4}{r^4}
    \right), \qquad
    g(r) = \frac{r^2}{\ell^2}, \qquad
    h(r) = \frac{1}{f(r)}.
\dle{eq:fg}
\ee
(In the limit $r_H \rightarrow 0$ this becomes pure AdS$_5$.)

More generally, by symmetry, a stack of non-rotating D3-branes
generates a metric of the form $ds_{10}^2 = ds_5^2 +
k(r)d\Omega_5^2$, where $ds_5$ is given in (\dr{5D})
\dc{Stelle:1996tz}.
In this case, since the metric coefficients are independent of the
angular coordinates $(\theta^1,\cdots,\theta^5)$, the unperturbed
brane dynamics are always characterized by a conserved angular
momentum around the S$^5$ \dc{Kehagias:1999vr}. As a result of the second
assumption above, we are thus effectively led to consider metrics
of the form (\dr{5D}): hence for the derivation of both the
unperturbed and perturbed equations of motion we keep $f,g,h$
arbitrary and consider the specific form (\dr{eq:fg}) only at the
end.

The embedding of the probe D3-brane
is given by $x^{\mu}=X^{\mu}(x^a)$.  (We have used
reparametrization invariance to choose the intrinsic worldsheet
coordinates $\sigma^a=x^a$.) For the unperturbed trajectory we
consider an infinitely straight brane parallel to the $x^a$
hyperplane but free to move along the $r$-direction:
\be
   X^{a} = x^{a}, \qquad X^{4} = R(t).
\dle{eq:position}
\ee
Later, in section \ref{sec:perturbed}, we will
consider a perturbed brane for which $X^{4} = R(t)+ \delta
R(t,\vec{x})$.

The induced metric on the brane is given by
\be \label{eq:pullbackmetric}
    \hat{\ga}_{ab} = g_{\mu \nu}(X)\frac{\dd X^{\mu}}{\dd x^a}
    \frac{\dd X^{\nu}}{\dd x^b}
\ee
(where the hat denotes a pullback), so that the line element on
the unperturbed brane worldsheet is
\be
    ds_4^2= \hat{\gamma}_{ab} dx^a dx^b
              = -(f(R)-h(R)\dot{R}^2)dt^2 + g(R) d\vec{x} \cdot d\vec{x}
              \equiv - d\tau^2 + a^2(\tau) d\vec{x} \cdot d\vec{x}.
\dle{eq:inducedmetric}
\ee
An observer on the brane therefore sees a homogeneous and
isotropic universe in which the time $\tau$ and the scale factor
$a(\tau)$ are given by
\be
  \tau = \int \sqrt{(f-h\dot{R}^2)}dt, \qquad a(\tau) =
  \sqrt{g(R(\tau))}.
\dle{eq:branetime}
\ee
The properties of the resulting Friedmann equation
depend on $f(R), g(R), h(R)$  (i.e.\ the bulk geometry) as well as
$\dot{R}$ (the brane dynamics) as discussed in
\dc{Kehagias:1999vr,Steer:2002ng} and summarized briefly below.


\subsection{Brane action and bulk 4-form field}
\label{subsec:braneaction}

In AdS$_5$-S, $B_{\mu\nu}$ vanishes, and we do not consider the
gauge field $F_{ab}$ on the brane. (For a detailed discussion of
the unperturbed brane dynamics with and without $F_{ab}$, which
essentially corresponds to radiation on the brane, see
\dc{Kehagias:1999vr,Steer:2002ng}.  Non-zero $B_{\mu\nu}$ has been
discussed in \dc{Youm:2000mi}.) Thus the brane action
(\dr{eq:dbiaction}) reduces to
\be
     S_{D3}=-T_3 \int d^{4}x \sqrt{-\hat{\ga}}
                -\rho_3 \int d^4 x \hat{C}_4
\dle{eq:dbiactionb}
\ee
where
\be
\hat{\ga}=\mbox{det}(\hat{\ga}_{ab}), \qquad
\hat{C}_4=C_{\mu\nu\si\rho}\frac{\dd X^{\mu}}{\dd x^0 }\frac{\dd
X^{\nu}}{\dd x^1}\frac{\dd X^{\si}}{\dd x^2}\frac{\dd
X^{\rho}}{\dd x^3}.
\dle{Chat}
\ee
and $C_{\mu\nu\si\rho}$ are components of the bulk RR 4-form
field.  The first term in (\dr{eq:dbiactionb}) is just the
Nambu-Goto action.

In the gauge (\dr{eq:position}), $\hat{\gamma}$ and $\hat{C}_4$
depend on $t$ only through $R$.  Thus rather than varying
(\dr{eq:dbiactionb}) with respect to $X^{\mu}$ and then
integrating the equations of motion, it is more straightforward to
obtain the equations of motion from the Lagrangian
\be
  {\cal{L}}= -\sqrt{-\hat{\ga}}-C \; \;
           = -\sqrt{fg^3-g^3h\dRR}-C
\dle{eq:lagrangian}
\ee
where $C=C(R) =
 \frac{\rho_3}{T_3}\hat{C}_4 =
q \hat{C}_4$. Since ${\cal L}$ does not explicitly depend on time,
the brane dynamics are parameterized by a (positive) conserved
energy $E = \frac{\dd {\cal{L}}}{\dd \dot{R}} \dot{R}-{\cal{L}}$
from which
\be
  \dRR=\frac{f}{h}\left(1-\frac{f g^3}{(E-C)^2}\right).
\dle{eq:rdotsquared}
\ee
Transforming to brane time $\tau$ defined in equation
(\ref{eq:branetime}) yields
\be
    R_{\tau}^2=\frac{(E-C)^2}{fg^3h}-\frac{1}{h} \dle{eq:retadotsquared}
\ee
where the subscript denotes a derivative with respect to $\tau$.

In order to analyze the brane dynamics in AdS$_5$-S where $f, g$
and $h$ are given in (\dr{eq:fg}), one must finally specify $C(R)$
or equivalently the 4-form potential $C_{\mu\nu\si\rho}$. To that
end\footnote{For the 10D AdS$_5$-S$\times$S$_5$ geometry the
solution for the 4-form field is given, for example, in
\dc{Kiritsis:1999tx}.  For completeness, we re-derive the result
starting directly from the 5D metric (\dr{eq:fg}).} recall  that
the 5D bulk action is
\be
  S=\frac{1}{2\kappa_5^2}\int d^5x \sqrt{-g}(R-2\Lambda)
  -\frac{1}{4\kappa_5^2}\int F_5 \wedge \ast F_5
\dle{eq:action}
\ee
where $\Lambda$ is the bulk cosmological constant
and  $F_5=dC_4$ is the 5-form field strength associated with the
4-form $C_4$.
The resulting
equations of motion are
\begin{eqnarray}
  R_{\mu\nu} &=& \frac{2}{3}\Lambda g_{\mu\nu}
     +\frac{1}{2 \cdot 4!}\left(F_{\mu \beta \gamma \delta \epsilon}
     {F_{\nu}}^{\beta \gamma \delta \epsilon}
     -\frac{4}{3\cdot5}F_{\alpha \beta \gamma \delta \epsilon}
     F^{\alpha \beta \gamma \delta \epsilon} g_{\mu\nu}\right), \dle{eq:ricci} \\
  d \ast F_5 &=&  \frac{1}{2}\frac{1}{\sqrt{fg^3h}}\left(
       \left(\frac{f'}{f}+3\frac{g'}{g}+\frac{h'}{h}\right)F_{01234}-
       2F_{01234}'\right)dr = 0
  \dle{eq:eomcterm}
\end{eqnarray}
where the prime denotes a derivative with respect to $r$.  In
AdS$_5$-S, $R_{\mu\nu}=-\frac{4}{\ell^2} g_{\mu\nu}$ and
Eq.~(\ref{eq:eomcterm}) gives
\be
  \frac{\ell^3}{r^3}\left(\frac{3}{r}F_{01234}-F_{01234}'\right)=
  0 \qquad \Longrightarrow \qquad F_{01234}= c \frac{r^3}{\ell^4}
\dle{eq:solF}
\ee
where $c$ is a dimensionless constant (see for example
\dc{Carter:2001nj}). (Note that this solution satisfies $dF_5=0$
since the only non-zero derivative is $\dd_4 F_{01234}$ which
vanishes on anti-symmetrizing.)
Integration gives
\be
  C_{0123}= v \frac{r^4}{\ell^4} + w
  \dle{eq:C0123}
\ee
where $v=c/4$ and $w$ are again dimensionless constants. Hence the
function $C(r)$ appearing in Eq. (\ref{eq:lagrangian}) is
\be
  C(r) =
  q C_{0123}=q v \frac{r^4}{\ell^4} + q w.
\dle{eq:C}
\ee

In 10 dimensions the constant $c$ (and hence $v$) is fixed by
$\int \ast F = Q$ ,
and $w$ may be determined by imposing (before taking the near
horizon limit -- hence with metric (\dr{eq:sugrametric})) that the
4-form potential should die off at infinity \dc{Kiritsis:1999tx}.
This second argument is not applicable here. Instead, we fix $v$
and $w$ in the following way: consider the motion of the
unperturbed brane  seen by a bulk observer with time coordinate
$t$. One can define an effective potential $V_{\rm eff}^{t}$
through
\be
    \frac{1}{2}\dRR + V_{\rm eff}^{t} \equiv E
  \dle{eq:kinetict}
\ee
so that on using equation (\ref{eq:rdotsquared}),
\be
    V_{\rm eff}^{t}(E,q,R)=
    E - \frac{1}{2}\left(\frac{R}{\ell}\right)^{4} {\alpha^2}
    \left[ 1 - \left(\frac{R}{\ell}\right)^8 \frac{\alpha}{(E - C)^2} \right]
\dle{eq:vefft}
\ee
(see Fig.\ref{fig2}) where
$$
\alpha  = 1- \frac{r_H^4}{R^4}
$$
and $C=C(R)$ is given in (\dr{eq:C}). We now use the fact that
there is no net force between static BPS objects of like charge,
and hence in this case the effective potential should be
identically zero. Here, such a configuration is characterized by
$r_H = 0, q=1, E=0$: imposing that $V_{\rm eff}^{t}=0$ for all
$R$, forces $v = \pm 1$ and, in this limit, $w = 0$. Second we
normalize the potential such that $V_{\rm eff}^{t}(E,q=1,R
\rightarrow \infty)=0$ for arbitrary values of the energy $E$ and
$r_H$. This leads to
\be v = - 1, \qquad w = + \frac{r_H^4}{2\ell^4}.
\dle{eq:abresult}
\ee
In particular for $E=0$, then the brane has zero kinetic energy at
infinity. Even in this case the potential is not flat, unless
$r_H=0$,  as can be see in Fig.\dr{fig1}. According to this
normalization
\be
    C(r)= - q\frac{r^4}{\ell^4} + q\frac{r_H^4}{2\ell^4}
    \dle{finalC}
\ee
as in the 10D case \dc{Kiritsis:1999tx}. Notice that since the
combination appearing in the equation of motion for $R$ is $E-C$,
the constant $w$ only acts to shift the energy. For later purpose
we define the shifted energy $\tilde{E}$ by
\be
\tilde{E} = E - q w = E - q \frac{r_H^4}{2 \ell^4}.
\ee


\begin{figure}[htbp]
\centerline{\epsfxsize=3.5in \epsfbox{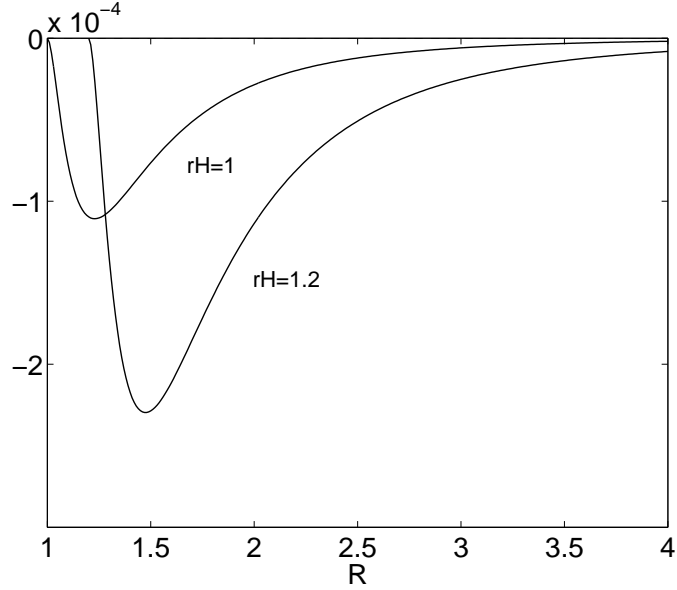}}
\caption{\label{fig1}  $V^{t}_{\rm eff}(E,q,R)$ for $E=0$, $q=1$,
$\ell=4$ and different values of $r_H$. For $R \rightarrow \infty$
the potential goes to zero according to our normalization. When
$r_H=0$, the potential is exactly flat.}
\end{figure}

Finally we comment that substitution  of (\ref{eq:solF}) into the
equation (\ref{eq:ricci}) determines the bulk
cosmological\footnote{Equivalently we could have started from the
10D SUGRA action, used the 10D solution for $F$ (which is
identical to (\dr{finalC})) and then integrated out over the
5-sphere.  After definition of the 5D Newton constant in terms of
the 10D one, the above cosmological constant term is indeed
obtained, coming from the 5-sphere Ricci scalar.} constant to be
given by $\ell^2 \Lambda = -6 - c^2/4 = - 10$.


\subsection{Brane dynamics and Friedmann equation}
\label{subsec:friedmann}

We now make some comments regarding the unperturbed motion of the
3-brane through the bulk, $R(\tau)$, as seen for an observer on
the brane.  This will be useful in section~\dr{sec:perturbed} when
discussing perturbations.  Recall that since $a(\tau) =
R(\tau)/\ell$ (see Eq.~(\dr{eq:branetime})), an `outgoing' brane
leads to cosmological expansion.  Contraction occurs when the
brane moves inwards. For the observer on the brane, one may define
an effective potential by
\be
    \frac{1}{2}{R_{\tau}}^2 + V_{\rm eff}^{\tau}=E
    \dle{eq:effpotentialdef}
\ee
whence, from Eq.~(\ref{eq:retadotsquared}),
\be
  V_{\rm eff}^{\tau}(E,q,R)= E + \frac{1}{2} \left( \frac{\ell}{R}
  \right)^6 \left[ \alpha \left(\frac{R}{\ell}\right)^8 - (E-C)^2 \right].
  \dle{eq:effpotential}
\ee

\begin{figure}[htbp]
\centerline{\epsfxsize=3.5in \epsfbox{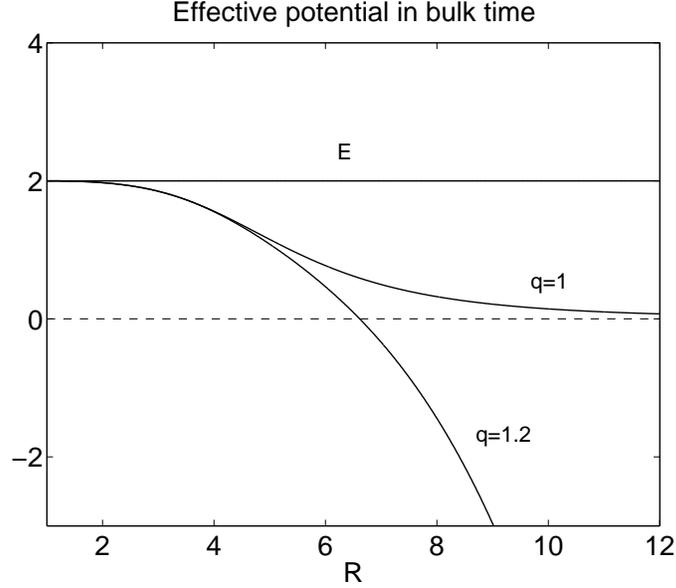}}
\caption{\label{fig2} $V^{t}_{\rm eff}(E,q,R)$ for $E=2$, $r_H=1$,
$\ell=4$. For a BPS-brane ($q=1$), $V_{\rm eff}^t \rightarrow 0$
as $R \rightarrow \infty$ according to our normalization. This
should be contrasted with a non-BPS brane e.g. with $q=1.2$. Note
that $V_{\rm eff}^t(E,q,R=r_H)=E$. Any inwardly moving
(contracting) brane takes an infinite amount of $t$-time to reach
the horizon. }
\end{figure}

\begin{figure}[htbp]
\centerline{\epsfxsize=3.5in \epsfbox{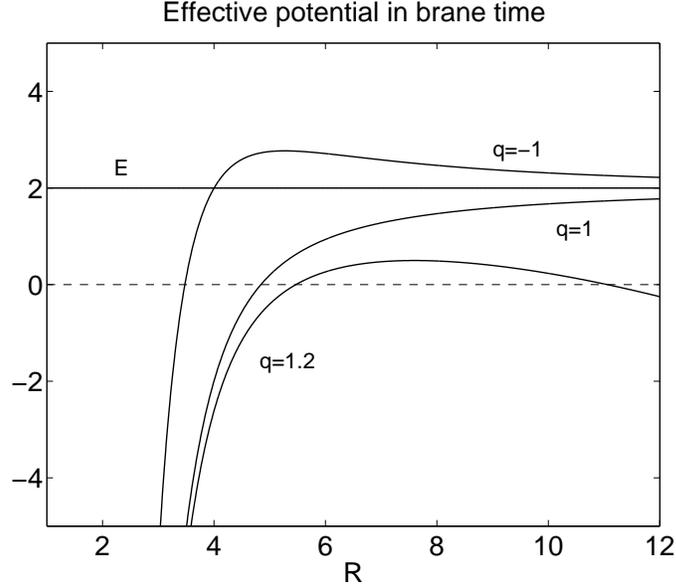}}
\caption{\label{fig3} $V^{\tau}_{\rm eff}(E,q,R)$ for the same
parameters as in Fig.\ref{fig2}.  A BPS brane has zero kinetic
energy at infinity corresponding to a vanishing cosmological
constant on the brane. Otherwise, the cosmological constant is
$\propto q^2-1$. A BPS anti-brane is allowed to move only in a
restricted range of $R$: after having reached a maximal scale
factor, the universe starts contracting. Any inwardly moving brane
falls into the black hole in a finite $\tau$.}
\end{figure}

Consider a BPS brane $q=+1$ (see Fig.\ref{fig3}). As noted above,
for $r_H= E = 0$ one has $V_{\rm eff}^{\tau} = 0$ so that the
potential is flat. For $r_H \neq 0$, $V_{\rm eff}^{\tau}$ contains
a term $\propto -R^{-6}$, and the probe brane accelerates towards
the horizon which is reached in finite ($\tau$-)time.  On the
other hand, for a bulk observer with time $t$, it takes infinite
time to reach the horizon where $V_{\rm eff}^{t} = E$, (see
Fig.~\ref{fig2}).

From equations (\dr{eq:retadotsquared}) and (\dr{eq:C}) it is
straightforward to derive a Friedmann-like equation for the brane
scale factor $a(\tau)$ \dc{Kehagias:1999vr,Steer:2002ng}:
\be
    H^2= \left(\frac{a_{\tau}}{a}\right)^2 =
    \frac{1}{\ell^2}\left[
    \frac{ \tE^2}{a^8} +
    \frac{1}{a^4} \left(2q\tE + \frac{r_H^4}{\ell^4}\right)
    + (q^2-1)\right].
    \dle{eq:friedmann}
\ee
The term in $1/a^8$ (a `dark fluid' with equation of state
$\tilde{p}=5/3\tilde{\rho}$) dominates at early times. The second
term, in $a^{-4}$, is a `dark radiation' term.  As discussed in
\dc{Steer:2002ng}, the part proportional to $r_H$ corresponds to
the familiar dark radiation term in conventional $Z_2$-symmetric
(junction condition) brane cosmology, where it is associated with
the projected bulk Weyl tensor.  When $\tilde{E}$ is non-zero,
$Z_2$-symmetry is broken\footnote{When making the link between
mirage cosmology and the junction condition approach, $\tE \propto
M_- - M_+$ where $M_{\pm}$ are the black-hole masses on each side
of the brane \dc{Steer:2002ng}.} \dc{Steer:2002ng} and this leads
to a further dark radiation term \dc{Carter:2001nj,Kraus:1999it}.
The last term in (\dr{eq:friedmann}) defines an effective
4-dimensional cosmological constant $\La_4 \equiv
\frac{1}{\ell^2}(q^2-1)$ which vanishes if the (anti) brane is BPS
(i.e.\ $q = \pm 1$). All these terms have previously been found
both in `mirage' cosmology and conventional brane cosmology
\dc{Steer:2002ng,Carter:2001nj}.

Notice that the dark radiation term above has a coefficient
\be
\mu \equiv 2q\tE + \frac{r_H^4}{\ell^4} = 2qE -
\frac{r_H^4}{\ell^4} (q^2 - 1)
\dle{coeffRm4}
\ee
which is positive for $q=+1$ (since $E \geq 0$). However, for BPS
anti-branes $q=-1$, the coefficient (\dr{coeffRm4}) is negative
unless $E=0$. Thus when $E \neq 0$ and $q=-1$ there is a regime of
$R$ for which $H^2$ is negative. In Fig.\ref{fig3} this is
represented by the forbidden region where the potential exceeds
the total energy $E$. At $V_{\rm eff}^{\tau}=E$ the Hubble
parameter is zero and an initially expanding brane starts
contracting. On the contrary, we do not obtain bouncing solutions
in our setup, regardless of the values of $q$ and $E$. Bouncing
and oscillatory universes are discussed  in e.g.
\dc{Mukherji:2002ft,Kachru:2002kx,BraxSteer:2002}.


The Friedmann equation (\ref{eq:friedmann}) can  be solved
exactly. In the BPS case, $\La_4 = 0$, the solution is
\begin{equation}
a(\tau)^4 = a_i^4 + \frac{4\mu}{\ell^2}(\tau-\tau_i)^2 \pm
\frac{4}{\ell}(\tau - \tau_i)(\tE^2 + \mu a_i^4)^{1/2}
\end{equation}
where $a_i$ is the value of the scale factor at the initial time
$\tau_i$, and the $\pm$ determines whether the brane is moving
radially inwards or outwards. In the next section when we solve
the perturbation equations, it will be sufficient to consider
regimes in which only one of the terms in equation
(\ref{eq:friedmann}) dominates. These will be given in section
\ref{sec:perturbed}.

One might wonder whether it is possible to obtain a term $\propto
a^{-3}$ (dust) in the Friedmann equation, and also one
corresponding to physical radiation on the brane (rather than dark
radiation).  Physical radiation comes from taking $F_{ab} \neq 0$
in (\dr{eq:dbiaction}) \dc{Kehagias:1999vr}, and a `dark' dust
term has been obtained in the non-BPS background studied in
\dc{Brax:2002qw}.  Finally, a curvature term $a^{-2}$ has been
obtained in \dc{Youm:2000ke}.


\section{Perturbed equations of motion}
\dle{sec:perturbed}

In this section we consider perturbations of the brane position
about the zeroth order solution $R(t)$ given in
(\dr{eq:rdotsquared}).  Once again we work with the metric
(\dr{5D}), specializing to AdS$_5$-S only at the end.  The
perturbed brane embedding $X^{4} = R(t)+ \delta R(t,\vec{x})$
leads to perturbations, $\delta \hat{\gamma}_{ab}$, of induced
metric on the brane and these are discussed in
section~\dr{sec:bardeen}. Note that these perturbations about the
flat homogenous and isotropic solution are not sourced by matter
on the brane, and their evolution will depend on the unperturbed
brane dynamics and hence on $E$. We now derive an equation for the
evolution of the perturbed brane and try to see if there are
instabilities in the system.

\subsection{The second order action}

Since we consider a codimension one brane, the fluctuations about
the unperturbed moving brane can be described by a single scalar
field $\phi(x^a)$ living on the unperturbed brane world sheet
\dc{Guven:1993ew}. To describe the dynamics of $\phi(x^a)$ (which
is defined below), we use the covariant formalism developed by
\dc{Guven:1993ew} to study perturbed Nambu-Goto walls. (For other
applications, see also \dc{Garriga:1991ts,Buonanno:1995bf}.) The
perturbed brane embedding is given by
\be
    X^{\mu}(t,\vec{x}) = \bar{X}^{\mu}(t) + \phi(t,\vec{x}) n^{\mu}(t)
\dle{eq:pertemb}
\ee
where $\bar{X}^{\mu}(t)$ is the unperturbed embedding, and
physical perturbations are only those transverse to the brane (see
also section \dr{sec:bardeen}). The unit spacelike normal to the
unperturbed brane, $n^{\mu}(t) = n^\mu(\bar{X}^{\mu}(t))$, is
defined through
\be
    g_{\mu\nu}n^\mu \frac{\dd \bar{X}^\nu}{\dd x^{a}}=0,
\qquad g_{\mu\nu}n^\mu n^\nu=1
\dle{eq:ortho}
\ee
so that
\be
    n^\mu=\left( \dR\sqrt{\frac{h}{f(f-h \dot{R}^2)}}, 0,0,0,
    \sqrt{\frac{f }{h(f-h \dot{R}^2)}}\right).
\dle{nn}
\ee
Thus for a 5D observer comoving with the brane, $\phi$ (which has
dimensions of length) is the measured deviation from the
background solution of the previous section \dc{Garriga:1991ts}.
For an observer living on the brane, the perturbations in the FRW
metric generated by $\phi$ are discussed in section
\dr{sec:bardeen} in terms of the gauge invariant scalar Bardeen
potentials.


An equation of motion for $\phi$ can be obtained by substituting
(\dr{eq:pertemb}) into the action (\dr{eq:dbiactionb}) and
expanding to second order in $\phi$. The terms linear in $\phi$
give the background (unperturbed) equations of motion studied in
the previous section
--- now we are interested in the terms quadratic in $\phi$ which give the
linearized equations of motion. A similar analysis was carried out
by Garriga and Vilenkin \dc{Garriga:1991ts} for Nambu-Goto cosmic
domain walls in Minkowski space and was generalized by Guven
\dc{Guven:1993ew} for arbitrary backgrounds.  For the action
(\dr{eq:dbiactionb}), the quadratic term is \dc{Buonanno:1995bf}
\be
    S_{\phi^2}=-\frac{1}{2}\int d^4 x
    \sqrt{-\hat{\ga}}\left[(\hat{\nabla}_{a}\phi)(\hat{\nabla}^{a}\phi)
    -\left({\hat{K}^{a}}_{b}{\hat{K}^{b}}_{a} + R_{\mu\nu}n^\mu
    n^\nu\right)\phi^2\right].
\dle{eq:phiaction}
\ee
Here $\hat{\nabla}$ is the covariant derivative with respect to
the induced metric $\hat{\gamma}_{ab}$, and the extrinsic
curvature tensor $\hat{K}_{ab}$ is given by
\be\dle{eq:excurv}
    \hat{K}_{ab}=(\nabla_\nu n_\mu) \frac{\dd \bar{X}^\mu}{\dd x^{a}}
\frac{\dd \bar{X}^\nu}{\dd x^{b}}
\ee
where $\nabla$ is the covariant derivative with respect to the 5D
metric $g_{\mu\nu}$.  Finally, $R_{\mu\nu}$ is the Ricci tensor of
the metric $g_{\mu\nu}$. Apart from $\phi$, all the terms in
(\dr{eq:phiaction}) are unperturbed quantities. Note that there is
no contribution to $S_{\phi^2}$ from the Wess-Zumino term of
action (\ref{eq:dbiactionb}):  all terms quadratic in $\phi$
cancel since $C_{0123}$ is the only non-zero component of the
4-form field.  However $C$ does enter into the term linear in
$\phi$ and hence into the background equations of motion, as
analyzed in the previous sections.

Variation of the action (\ref{eq:phiaction}) with respect to $\phi$ leads to
the equation of motion
\be
    \hat{\nabla}^{a}\hat{\nabla}_{a}\phi+[{\hat{K}^{a}}_{b}
    {\hat{K}^{b}}_{a} + R_{\mu\nu}n^\mu n^\nu]\phi = 0
    \dle{pertf}
\ee
or equivalently
\be
\hat{\nabla}^{a}\hat{\nabla}_{a}\phi - m^2 \phi =
    0
\dle{pert}
\ee
where
\be
m^2 = - [{\hat{K}^{a}}_{b}
    {\hat{K}^{b}}_{a} + R_{\mu\nu}n^\mu n^\nu].
\dle{m2def}
\ee

To determine the extrinsic curvature contribution to (\dr{m2def}),
it is simpler to calculate first the five dimensional extrinsic
tensor defined by
\be
    K^{\mu}_{\nu}=\ga^{\la \mu} \nabla_{\la} n_{\nu}
\ee
where $\ga^{\la \mu}=g^{\la \mu}-n^{\la} n^{\mu}$ and then use
$$
{\hat{K}^{a}}_{b} {\hat{K}^{b}}_{a}=
     K^{\mu}_{\nu} K^{\nu}_{\mu}.
$$
On defining $T$ by
$$
T \equiv \left(\frac{d\tau}{dt}\right)^2 =
f-h\dRR=\frac{f^2g^3}{(E-C)^2},
$$
the non-zero components of $K^{\mu}_{\nu}$ are
\begin{eqnarray}
    K^0_0&=&\frac{1}{T^{5/2}}f^{3/2}h^{1/2}
    \left(\ddot{R}-\frac{f'}{f}\dRR +\frac{1}{2}\frac{h'}{h}\dRR
    + \frac{1}{2}\frac{f'}{h}\right),\\
    K^0_4&=&-\frac{h \dR}{f} K^0_0,\\
    K^1_1&=&\frac{1}{T^{1/2}}\left(\frac{f}{h}\right)^{1/2}\frac{1}{2}\frac{g'}{g} = K^2_2=K^3_3,\\
    K^4_4&=&-\frac{h \dRR}{f}K^0_0
\end{eqnarray}
so that
\be
     {\hat{K}^{a}}_{b} {\hat{K}^{b}}_{a}=
      \frac{1}{T}\frac{f}{h}\left(3\left(\frac{g'}
{g}\right)^2+3\frac{g'}{g}\frac{C'}{E-C}
        +\left(\frac{C'}{E-C}\right)^2\right).
\ee
The Ricci term is
\ba
     R_{\mu\nu}n^\mu n^\nu&=&
     -\frac{1}{4h}\left(2\frac{f''}{f}-\left(\frac{f'}{f}\right)^2
        +3\frac{f'}{f}\frac{g'}{g}-\frac{f'}{f}\frac{h'}{h} \right)
\\
\nn
    &+& \frac{3}{4}\frac{1}{T}\frac{f}{h}\left(
         \frac{f'}{f}\frac{g'}{g}-2\frac{g''}{g}+\left(\frac{g'}{g}
         \right)^2 + \frac{g'}{g}\frac{h'}{h}\right).
\nn
\ea
Collecting these results gives
\begin{eqnarray}\label{eq:effmass}
  m^2 &=& -\frac{3}{4}\frac{(E-C)^2}{fg^3h}
          \left(\frac{f'}{f}\frac{g'}{g}-2\frac{g''}{g}+5\left(\frac{g'}{g}\right)^2
          +\frac{g'}{g}\frac{h'}{h}
       +4\frac{g'}{g}\frac{C'}{E-C}+\frac{4}{3}
      \left(\frac{C'}{E-C}\right)^2\right)
\nn
\\
      &+&\frac{1}{4h}\left(2\frac{f''}{f}-\left(\frac{f'}{f}\right)^2 +3\frac{f'}{f}\frac{g'}{g}
      -\frac{f'}{f}\frac{h'}{h}\right).\nn
\end{eqnarray}
In the remainder of this section we try to obtain approximate
solutions for $\phi$ from equation (\dr{pert}).  Some aspects of
this calculation are clearer in brane time $\tau$, and others in
conformal time $\eta$ (where $\eta=\int d\tau/a(\tau)$).  Of
course the results are independent of coordinate system. For these
reasons we have decided to present both approaches beginning with
brane time.


\subsection{Evolution of perturbations in brane time $\tau$}
\label{subsec:branetime}

On using the definition of brane time $\tau$ in equation
(\dr{eq:branetime}), the kinetic term in (\dr{pertf}) is given by
\ba
         \hat{\nabla}^{a}\hat{\nabla}_{a}\phi &=&
-  \phi_{\tau \tau} - 3 H \phi_{\tau}
  + \frac{1}{a^2}[\phi_{x^1x^1}+\phi_{x^2x^2}+\phi_{x^3x^3}].
\nn
\ea
(In conformal time the factor of $a^{-2}$ multiplying the spacial
derivatives disappears --- see below.) We now change variables to
$ \vph = a^{3/2} \phi$ so that (\dr{pert}) becomes
\be
    \varphi_{\tau \tau}
  - \frac{1}{a^2}[\varphi_{x^1x^1}+\varphi_{x^2x^2}+\varphi_{x^3x^3}]
  + M^2(\tau)\varphi = 0
\label{eq:perttau}
\ee
where
\ba
  M^2(\tau)  &=& m^2 - \frac{3}{4} \left[ \left(\frac{a_{\tau}}{a}\right)^2 + 2
  \frac{a_{\tau\tau}}{a} \right]
\nn
\\
&=& m^2 - \frac{3}{4} \left[\frac{g''}{g}  R_{\tau}^2 -
\frac{1}{4} \left(\frac{g'}{g} \right)^2 R_{\tau}^2 + \frac{g'}{g}
R_{\tau\tau} \right]
\\
&=&   \frac{3}{4}\frac{(E-C)^2}{fg^3h}
          \left(-\frac{1}{2}\frac{f'}{f}\frac{g'}{g}
          +\frac{g''}{g}-\frac{13}{4}\left(\frac{g'}{g}\right)^2
           -\frac{1}{2}\frac{g'}{g}\frac{h'}{h}
       -3\frac{g'}{g}\frac{C'}{E-C}-\frac{4}{3}\left(\frac{C'}{E-C}\right)^2\right)
\nn
\\
  &+&\frac{1}{4h}\left(2\frac{f''}{f}-\left(\frac{f'}{f}
\right)^2+3\frac{f'}{f}\frac{g'}{g}
      -\frac{f'}{f}\frac{h'}{h}+3\frac{g''}{g}-
\frac{3}{4}\left(\frac{g'}{g}\right)^2
      -\frac{3}{2}\frac{g'}{g}\frac{h'}{h}\right).
\dle{meta}
\end{eqnarray}
This expression is valid for any $f$, $g$ and $h$. We now
specialize to AdS$_5$-S in which case
\ba
  M^2(\tau)&=&
\frac{1}{\ell^2} \left[ -\frac{33}{4}\frac{ \tE^2}{a^8}
  +\frac{3}{4}\frac{1}{a^4}\left(2q\tE+\frac{r_H^4}{\ell^4}\right)
  -\frac{25}{4}(q^2-1) \right]
\nn
\\
  & = & -\frac{33}{4}H^2
  +\frac{9}{a^4\ell^2}\left(2q\tE+\frac{r_H^4}{\ell^4}\right)
  +2\frac{q^2-1}{\ell^2} .
\dle{eq:effmassadsss}
\ea
Notice that there are regimes of $a$ in which $M^2 < 0$ --- such
as, for instance, for small $a$ where the $a^{-8}$ term dominates
--- and furthermore that the location of these regimes depends on
the energy $E$ of the brane. We also see that since $M^2 \sim
H^2$, instabilities will occur for modes with a wavelength greater
than $H^{-1}$. Figure \dr{fig4} shows the typical shape of $M^2$
as a function of $a$ for fixed energy and different $q$. In the
following, we only discuss  cases with $q^2 \geq 1$ as the 4D
cosmological constant is positive.

\begin{figure}[htbp]
\centerline{\epsfxsize=3.5in \epsfbox{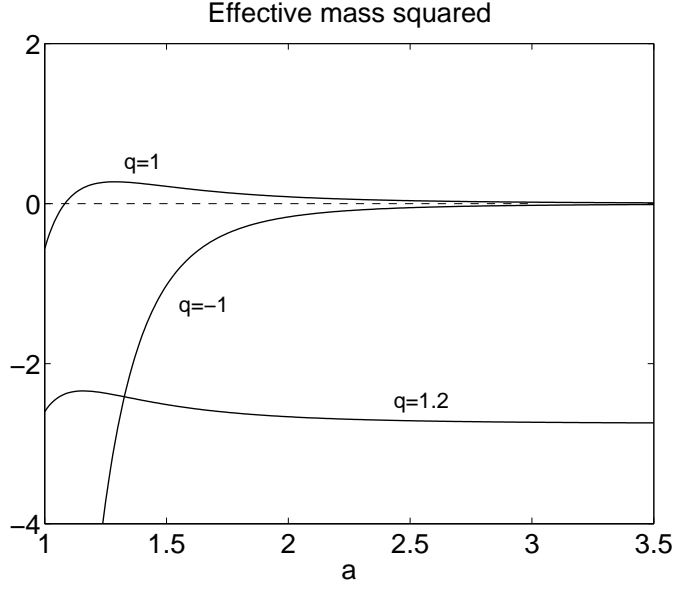}}
\caption{\label{fig4}The dimensionless quantity $M^2 \ell^2$ as a
function of $a$ for $E=1, \ell=1, r_H=1$. Here, the effective mass
squared is positive in a certain range only for the BPS-brane.
Note that the negative $M^2 \ell^2$ region is not hidden behind
the horizon. }
\end{figure}

Analysis of equation (\ref{eq:perttau}) is simpler in Fourier
space where
\begin{equation}
    \vpk(\tau) =
    \int d^3 x
    \varphi(\tau,\vec{x})e^{-i\vec{k}\cdot \vec{x}}
\label{eq:fourier}
\end{equation}
and $k$ is a comoving wave number related to the physical wave
number $k_p$ by $k = a k_p$.  Thus (\ref{eq:perttau}) becomes
\be
\vpkt + \frac{1}{a^2}\left(k^2 - k_c^2(\tau) \right) \vpk = 0
\dle{criteq}
\ee
where the time dependent critical wave number, $k_c^2(\tau)$, is
given by
\be
    k_c^2(\tau) = -M^2(\tau)a^2.
\dle{kcdef}
\ee

One might suppose that for $M^2> 0$ all modes are stable. However,
due to the $\tau$-dependence of $k_c$ this is not necessarily true
(as we shall see in equation (\ref{eq:surprise})).


Our aim now is to determine the $a$-dependence of $\vpk$.  We
proceed in the following way: notice first that the Friedmann
equation (\dr{eq:friedmann}) and the expression for $M^2(\tau)$ in
(\dr{eq:effmassadsss}) both contain terms in $a^{-8}$, $a^{-4}$
and $a^0$. We will focus on a regime in which one of these terms
dominates. Then the Friedmann equation can be solved for $a(\tau)$
which, on substitution into (\dr{eq:effmassadsss}), gives
$M^2(\tau)$. A final substitution of $M^2(\tau)$ into the
perturbation equation (\ref{criteq}) for $\vpk$ enables this
equation to be solved in each regime. We consider the following
cases: {\it i}) $q=+ 1$, {\it ii}) $q =- 1$ and {\it iii}) $q^2 >
1$.

\subsubsection{BPS brane: $q= + 1$}

For a BPS brane, the Friedmann equation (\dr{eq:friedmann}) and
effective mass $M^2(\tau)$ are given by
\ba
H^2 &=& \frac{1}{\ell^2} \left[ \frac{\tilde{E}^2}{a^8} +
\frac{2 E}{a^4} \right],
\dle{H2E0}
\\
M^2(\tau) &=& \frac{1}{\ell^2} \left[ -\frac{33}{4} \frac{\tilde{E}^2}{a^8} +
\frac{3}{2} \frac{E}{a^4} \right] .
\dle{M2E0}
\ea
The $E$-dependence of these equations slightly complicates the analysis
of these equations, and hence we begin with the simplest case in which $E=0$.
\\
\\
{\bf Case 1}: $E=0$

When $E=0$ --- the static limit in which the probe has zero
kinetic energy at infinity (see Fig.\dr{fig1}) --- only the term
proportional to $a^{-8}$ survives in (\dr{H2E0}) and (\dr{M2E0}).
Recall
that when $r_H$ vanishes the potential $V^{\tau}_{\rm{eff}}$ is
flat. Furthermore, since $\tE \propto r_H^4 = 0$, it follows from
(\dr{M2E0}) that $M^2(\tau)=0$ in this limit: as expected, a BPS
probe brane with zero energy  in AdS$_5$ has no dynamics and is
completely stable.

When $r_H \neq 0$, $M^2(\tau) < 0 \; \forall  \tau$,  and the
solution of (\dr{H2E0}) is
\be
a(\tau)^4 = a_i^4 \pm \frac{2a_H^4}{\ell} (\tau - \tau_i).
\dle{Rsoln}
\ee
Here $a_i \geq a_H \equiv r_H/\ell$ is the initial position of the
brane at $\tau=\tau_i$, and the choice of sign determines whether
the brane is moving radially inwards ($-$) or outwards (+): this
is a question of initial conditions. Let $R_h = 1/|Ha|$ denote the
(comoving) Hubble radius. Then it follows from (\dr{M2E0}) and the
definition of $k_c^2$ in (\dr{kcdef}) that
\be
\frac{1}{\lambda_c} \sim |k_c(\tau)| \sim |H a| = \frac{1}{R_h}
\ee
where we neglect numerical factors of order 1. Thus the critical
wave length is $\lambda_c \sim R_h$. (Notice that $R_h$ is minimal
at $a_H$ and increases with $a$.)
\\

For {\em superhorizon} modes $\lambda \gg R_h$ or $|k| \ll |k_c|$,
and in this limit the perturbation equation (\dr{criteq}) becomes
\be
\varphi_{k,\tau \tau}-\frac{k_c^2(\tau)}{a^2} \varphi_k = 0.
\ee
On inserting solution (\ref{Rsoln}) into $k_c^2$ one obtains
\be \label{eq:superhorizonsol}
\phi_k = \frac{\varphi_k}{a^{3/2}}= A_k a^4 + B_k a^{-3}
\ee
(where the constants $A_k$ and $B_k$ are determined by the initial
conditions).  Hence if the brane moves radially outwards the
superhorizon modes grow as $a^4 \propto \tau$.  If the brane is
contracting they grow $a^{-3}$.  In the near extremal limit,
($r_H \ll \ell$ or) $a_H \ll 1$, the amplitude of these
superhorizon modes can become very large suggesting that they are
unstable. Of course our linear analysis will break down when
$\phi$ becomes too large.
\\

Consider now {\em subhorizon} modes $\lambda \ll R_h$ or $|k| \gg
|k_c|$. Then (\dr{criteq}) is just $\vpkt + \left(k^2/a^2\right)
\vpk = 0$. However, in this case it is much easier to solve the
equation in conformal time $\eta$ where the factor of $a^{-2}$ is
no longer present. We anticipate the result from section
\dr{subsec:conformaltime}: it is
\be
\dle{eq:oscsol}
\phi_k =A_k\frac{e^{ik\eta}}{a}+B_k\frac{e^{-ik\eta}}{a}.
\ee
For an outgoing brane $a$ increases and subhorizon modes are
stable. For an ingoing brane $a$ decreases, and the amplitude of
the perturbation becomes very large in the near extremal limit.
(Note that as the brane expands, superhorizon modes eventually
become subhorizon, and similarly, on a contracting brane,
subhorizon modes become superhorizon.)

To conclude, when $r_H \neq 0$, $E=0$ and the brane expands,
superhorizon modes are unstable  whilst subhorizon modes are
stable. For a contracting brane, and in the near extremal limit,
both super- and sub-horizon modes are unstable.
\\
\\
{\bf Case 2: $E \neq 0$}

When the energy of the brane is non-zero the situation is more
complicated.  Notice first from (\dr{M2E0}) that $M^2(\tau)$ has
one zero at $a = a_c$ given by
\be \label{eq:zeroofm2}
 {a}_c^4  = \frac{11  \tilde{E}^2}{2E}.
\ee
Hence $M^2(\tau)$ is negative when $a < a_c$ and positive for $a >
a_c$ (see Fig.\dr{fig5}). However, since $a_c$ is $E$-dependent,
there may be ranges of $E$ for which the negative mass region is
hidden within the black hole horizon.  Indeed we find
\be
 a_c \leq a_H \qquad  \Longleftrightarrow \qquad E_- \leq E \leq E_+
\ee
where
\be \label{eq:Eplusminus}
E_{\pm} = \frac{a_H^4}{22} (13 \pm 4 \sqrt{3}).
\ee
The situation is shown schematically in Fig.\dr{fig5}.

\begin{figure}[htbp]
\centerline{\epsfxsize=3.5in \epsfbox{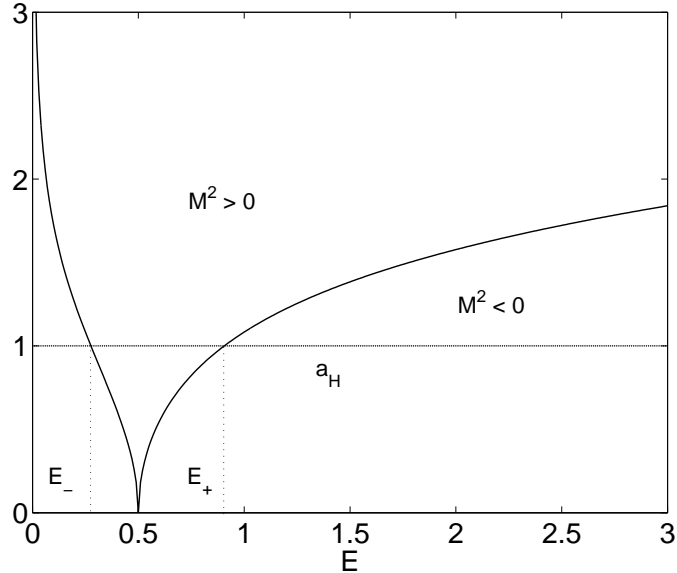}}
\caption{\label{fig5} The curve represents $a_c$, the zero of
$M^2(\tau)$, as a function of the energy $E$ as given in Eq.
\ref{eq:zeroofm2}. Below the curve the effective mass squared is
negative, above it is positive. For $E<E_-$ and $E>E_+$ the
$M^2(\tau)$ becomes negative already outside the horizon, whereas
for energies within the interval $E_-,E_+$ the $M^2(\tau)<0$
region is hidden within the horizon. The parameters chosen are
$q=1$,$r_H=1$ and $\ell=1$.}
\end{figure}


Now consider $H^2$ given in Eq. (\dr{H2E0}). The two terms are of
equal magnitude when $a=\tilde{a}_c = (\tilde{E}^2/{2E})^{1/4}
\sim a_c$.  Thus when $a \ll a_c$ (and hence in the regions in
which $M^2 < 0$ in Fig.\dr{fig5}), the dominant term in $H^2$ is
the one proportional to $a^{-8}$.  The system is therefore
analogous to the one considered above when $E=0$, and for
superhorizon modes the solution is given in
(\dr{eq:superhorizonsol}): for an outgoing brane $\phi_{k} \sim
a^4$. When $E\gsim E_+$ or $E \lsim E_-$, these regimes extend
down to the blackhole horizon: thus in the near extremal limit the
contracting brane will again be unstable since $\phi_{k} \sim
a^{-3}$.
\\

When $a \gg a_c$ (and hence in the regimes in which $M^2 > 0$ in
Fig.\dr{fig5}), the dominant term in $H^2$ is $\propto a^{-4}$ so
that
\be
a(\tau)^2 = a_i^2 \pm 2 \frac{\sqrt{2E}}{\ell} (\tau - \tau_i)
\dle{Rsoln2}
\ee
and
\be
k_c^2(\tau)= - M^2(\tau)a^2 = -\frac{3}{2} \frac{ E}{\ell^2 a^2}
\ee
On {\em superhorizon} scales the mode equation is
\be
\varphi_{k,\tau \tau} + \frac{3}{2} \frac{ E}{\ell^2 a^4}
\varphi_k = 0.
\dle{criky}
\ee
At first sight one might expect the solution to this equation to
be stable since $M^2 > 0$.  However, surprisingly, it is not.
(Indeed, below we will see that in conformal time the effective
mass is actually negative in this regime.) A change of variables
to $u=a^2$
shows that the solution of (\ref{criky}) is
\be \label{eq:surprise}
\varphi_k = A_k a^{3/2} + B_k a^{1/2}
\ee
which grows as $\tau^{3/4}, \tau^{1/4}$ respectively. Finally
\be
\phi_k =   A_k + B_k a^{-1}.
\dle{solnc}
\ee
For $E$ within the band $E_- \lsim E \lsim E_+$, the solution
(\dr{solnc}) for the modes is valid for all $a$ so that {\em
superhorizon} modes grow as $a^{-1}$ as the brane approaches the
black hole horizon.

When $E \gsim E_+$ or $E \lsim E_-$ these solutions are valid for
$a \gg a_c$. Thus for an expanding brane $\phi_k$ tends to a
constant value.
For a contracting brane, the term $\propto a^{-1}$ could become
important, though for small enough $a$ the relevant regime is that
considered above in which case the solution is given by
(\dr{eq:superhorizonsol}) and the superhorizon modes grow as
$a^{-3}$.

For {\em subhorizon} modes, the solution is still as given in
(\dr{eq:oscsol}).

\subsubsection{BPS anti-branes: $q = - 1$}
\dle{sub:anti}

Now the Friedmann equation (\dr{eq:friedmann}) and effective mass $M^2(\tau)$
become
\ba
H^2 &=& \frac{1}{\ell^2}\left[ \frac{\tilde{E}^2}{a^8} - \frac{2
E}{a^4} \right],
\dle{H2E0b}
\\
M^2(\tau) &=& -\frac{1}{\ell^2} \left[ \frac{33}{4} \frac{
\tilde{E}^2}{a^8} + \frac{3}{2} \frac{ E}{a^4} \right]
\dle{M2E0b}
\ea
so that $M^2$ is always negative, independently of $E$.  Note that
$H^2 > 0$ for $a <\tilde{a}_c$ where $\tilde{a}_c =
(\tE^2/2E)^{1/4}$.  However, since $\tE = E$ {\bf +} $a_H^4/2$ for
anti-branes, it follows that $\tilde{a}_c \geq a_H$ for {\it all}
$E$ (i.e.\ there are no energy bands to consider in the case of
anti-branes).  When $a \ll \tilde{a}_c$, $H^2 \propto M^2 \propto
a^{-8}$ and once again this is analogous to the case studied above
for $E=0$: {\em superhorizon} modes grow as $a^4$, and in the near
extremal limit the {\em subhorizon} modes on an ingoing brane are
unstable.


\subsubsection{Non-BPS branes: $q \neq \pm 1$}
\dle{sub:nonbps}

Here we shall only briefly discuss the case $q^2 > 1$ for large
$a$. Now, independently of $E$, there is a cosmological constant
dominated regime  (see Eq. (\dr{eq:friedmann})). There the
solution for the scale factor is
\be
a(\tau) = a(\tau_i) e^{\pm \sqrt{\Lambda_4} (\tau-\tau_i)}  \qquad
{\rm where} \qquad \Lambda_4 \equiv \frac{q^2-1}{\ell^2}.
\ee
In this regime, however, $M^2$ is negative with
\be
M^2(\tau) = - \frac{25}{4} \Lambda_4
\ee
and $R_h = \frac{1}{|Ha|}={\Lambda_4}^{-1/2}a^{-1}$.

For {\em subhorizon} modes ($\lambda \ll R_h$) the solution for
$\vpk$ is again given by (\dr{eq:oscsol}).  For {\em superhorizon}
modes, and considering an outgoing brane, there is an
exponentially growing unstable mode
\be
\pk = A_k e^{\sqrt{\Lambda_4} (\tau - \tau_i)} = A_k a.
\ee
Hence, this non-BPS brane is unstable for large $a$. It is not
clear to us why the acceleration due to the positive cosmological
constant does not rather stretch the perturbations away.


\subsection{Comments on an analysis in conformal time $\eta$}
\label{subsec:conformaltime}

It is instructive to carry out a similar analysis in conformal
time rather than brane time,
and we comment briefly on it here. In conformal time and
transformed to  Fourier space, Eq. (\dr{pert}) becomes
\be
    \phi_{k,\eta \eta} + 2 {\cal{H}} \phi_{k,\eta}
    +(k^2 + a^2 m^2) \phi_k =0
\dle{eq:phieta}
\ee
where ${\cal{H}}= a H$. The friction term can be eliminated by a
change of variables to $\psi = a \phi$, and the above equation
becomes
\be
\psi_{k,\eta \eta} + \left( k^2 - k_c^2(\eta) \right) \psi_k = 0
\dle{perteta}
\ee
where
$$
k_c^2(\eta) = -{\cal{M}}^2(\eta).
$$
and
\ba
  {\cal{M}}^2(\eta)  &=& a^2 m^2 - {a_{\eta \eta}}/{a}
\\
&=& g m^2 + \frac{1}{2} \left[ - \frac{g''}{g}  R_{\eta}^2
  + \frac{1}{2} \left(\frac{g'}{g} \right)^2 R_{\eta}^2 - \frac{g'}{g}
  R_{\eta\eta} \right]
\nn
\\
&=& -  \frac{(E-C)^2}{fg^2h}
          \left(
          \frac{1}{2} \frac{f'}{f}\frac{g'}{g}
          - \frac{g''}{g}+3\left(\frac{g'}{g}\right)^2
          + \frac{1}{2}\frac{g'}{g}\frac{h'}{h}
          + \frac{5}{2}\frac{g'}{g}\frac{C'}{E-C}
          + \left(\frac{C'}{E-C}\right)^2
          \right)
\nn
\\
  &+&\frac{g}{2h}
  \left(
  \frac{f''}{f}
  -\frac{1}{2}\left(\frac{f'}{f} \right)^2
  +\frac{3}{2}\frac{f'}{f}\frac{g'}{g}
  - \frac{1}{2}\frac{f'}{f}\frac{h'}{h}
  + \frac{g''}{g}
  - \frac{1}{2}\frac{g'}{g}\frac{h'}{h}\right).
\dle{eq:masseta}
\end{eqnarray}
Specializing to AdS$_5$-S yields
\be
 {\cal{M}}^2(\eta) = -\frac{1}{\ell^2}\left[\frac{10\tE^2}{a^6}
 +6(q^2-1)a^2 \right].
\dle{meta2}
\ee

Notice that in conformal time and for $|q| \geq 1$, $\cM^2(\eta)$
is {\em always} negative {independently of} $E$. From this, one
can  immediately see the instability for small $k$ in Eq.
(\ref{eq:surprise}), even though $M^2(\tau)$ can be positive in
that case. It is clear that the results on brane (in)stability
must be independent of whether or not the analysis is carried out
$\eta$ or $\tau$ time. We will see that this is indeed the case:
the reason is that not only the sign of the effective mass
squared, but also its functional dependence on time determines the
stability properties. We now summarize briefly some of the aspects
which differ between the $\tau$ and $\eta$ analysis.
\\

Consider the simplest case: $q=+1$ and $E=0$.  The solution of the
(conformal time) Friedmann equation is $a^3 = a_i^3 \pm 3
a_H^2(\eta-\eta_i)/2\ell$, and $k_c(\eta) \sim  |\cH| = 1/R_h$.
For {\em superhorizon} modes, $|k| \ll |k_c|$, Eq. (\dr{perteta})
reduces to $\psi_{k,\eta \eta} - k_c^2(\eta) \psi_k = 0$. Given
$a(\eta)$ and hence $k_c(a(\eta))$ it is straightforward to find
the solution which is, as expected, exactly that given in
(\dr{eq:superhorizonsol}). For {\em subhorizon} modes, $|k| \gg
|k_c|$, the solution was given in (\dr{eq:oscsol}).
\\

Consider now $q=+1, E>0$. Recall that in the $\tau$-time analysis
both $M^2(\tau)$ and $H^2$ contained terms in $a^{-4}$ and
$a^{-8}$ and, in particular, there was a regime in which
$M^2(\tau)$ was positive and proportional to $a^{-4} \propto H^2$.
In $\eta$-time, however, ${\cal{H}} \propto a^{-6} + a^{-2}$ with
${\cal{M}}^2$ is always being negative, $\propto - a^{-6}$.  Thus
whilst the $a \ll a_c$ regime reduces to that discussed above for
$E=0$, the $a\gg a_c$ regime is a little less clear.  There $\cH^2
\sim a(\eta)^{-2}$, but $\cM^2 \sim -a(\eta)^{-6}$. Thus
\be
a(\eta) = a_i \pm \frac{\sqrt{2E}}{\ell}(\eta - \eta_i)
\ee
and
\be
    k_c(\eta)^2 = - \cM(\eta)^2 = \frac{10 \tE^2}{\ell^2 a^6}.
\ee
Now $|k_c(\eta)| \sim |\cH|^3 \ell^2 = \ell^2/R_h^3$, and so one
can no longer identify the critical wavelength with the Hubble
radius.
For $|k| \ll |k_c|$  Eq. (\dr{perteta}) reduces to $d^2\psi_k/da^2
- (5\tE^2/E) (\psi_k/a^6) = 0$.  The solution is expressed in
terms of Bessel functions which, however, show exactly the same
behavior as (\dr{solnc}): namely $\phi_k=\psi_k/a$ tends to a
constant as $a \rightarrow \infty$. The other limit $a \rightarrow
0$ is not relevant as the above equation is only valid for $a \gg
{a}_c$.

We do not discuss further the case of $q=-1$ and $q\neq 1$ since
the results obtained in this approach are exactly as discussed in
sections \dr{sub:anti} and \dr{sub:nonbps}.


\section{Bardeen potentials}
\dle{sec:bardeen}

So far we have discussed the evolution of $\phi$, the magnitude of
the brane perturbation as seen by a 5D observer comoving with the
brane.  For an observer living {\it on} the brane, the perturbed
brane embedding gives rise to perturbations about the FRW
geometry.  Recall (see Eq. (\dr{eq:inducedmetric})) that for the
unperturbed brane
\ba
    d\overline{s}_4^2 &=& \hat{\bar{\gamma}}_{ab} dx^a dx^b
    \nn
    \\
    &=& -(f(R)-h(R)\dot{R}^2)dt^2 + g(R) d\vec{x} \cdot d\vec{x}
    \nn
    \\
    &\equiv&  -n^2(t) dt^2 + a^2(t) d\vec{x} \cdot d\vec{x}
    \dle{unp1}
\ea
where the bar on $\bar{\gamma}$ denotes that it is an unperturbed
quantity.  Note that the scale factors $n^2(t)$ and $a^2(t)$ pick
up their time-dependence through $R(t)$ --- for instance $a^2(t) =
g(R(t))$.  In this section we calculate $\delta
\hat{\gamma}_{ab}$ resulting from the perturbed embedding
(\dr{eq:pertemb}) and relate it to the Bardeen potentials.

Initially, rather than using the covariant form (\dr{eq:pertemb}),
let us write more generally
\ba
X^{0}(t,\vec{x}) &=& t + \zeta^0(t,\vec{x}),
\dle{X0}
\\
X^{i}(t,\vec{x}) &=& x^{i} + \zeta^{i}(t,\vec{x}),
\dle{Xi}
\\
X^4(t,\vec{x}) &=& R(t) + \epsilon(t,\vec{x}).
\dle{X1}
\ea
Below we will see that the perturbations $\zeta^{i}$ do not enter
into the two scalar Bardeen potentials which correspond to the two
degrees of freedom $\zeta^0$ and $\epsilon$. This is expected
since perturbations parallel to the brane are not physical and can
be removed by a coordinate transformation \dc{Deruelle:2000yj}.
Then only right at the end will we set $\zeta^0/n^0 = \epsilon/n^5
= \phi$.  We will find that the two Bardeen potentials are
proportional to each other and to $\phi$.

By definition, the perturbed brane embedding is given by
\ba
\hat{\gamma}_{ab} &\equiv& \hat{\bar{\gamma}}_{ab} + \delta
\hat{\gamma}_{ab}
\nn
\\
& = & g_{\mu \nu}(\bar{X} + \delta{X}) \frac{\partial}{\partial
x_a} (\bar{X}^{\mu} + \delta{X}^{\mu}) \frac{\partial}{\partial
x_b} (\bar{X}^{\nu} + \delta{X}^{\nu}) .
\dle{pertg}
\ea
Evaluating $\delta \hat{\gamma}_{ab}$ to first order for the
perturbed embedding (\dr{X0})-(\dr{X1}) and the general bulk
metric (\dr{5D}), one obtains
\ba
\delta \hat{\gamma}_{00} & = & \epsilon (-f'+ h' \dot{R}^2) + 2( -
\dot{\zeta}^0 f +  \dot{\epsilon} h  \dot{R} ),
\dle{pert00}
\\
\delta \hat{\gamma}_{0i} & = & - (\partial_i \zeta^0) f +
\dot{\zeta}^{i} g + (\partial_i \epsilon) h \dot{R},
\dle{pert0i}
\\
\delta \hat{\gamma}_{ij} & = &  \epsilon g'\delta_{ij}+
(\partial_i \zeta_j + \partial_j \zeta_i )g.
\dle{pertij}
\ea
Note the terms proportional to $\epsilon$ come from the Taylor
expansion of $g_{\mu \nu}(\bar{X} + \delta{X})$ in (\dr{pertg}) to
first order.

In the usual way, the perturbed line element on the brane is
written as
\be
    ds_4^2 = -n^2 ( 1 + 2 \underline{A}) dt^2 - 2 a n \underline{B}_i
dt dx^{i} + a^2(\delta_{ij} + \underline{h}_{ij}) dx^i dx^j
\ee
where $n(t)$ and $a(t)$ are defined in (\dr{unp1}), and as usual
vectors are decomposed into a scalar part and a divergenceless
vector component e.g.\
\be
\underline{B}_i = \partial_i \underline{B} +
\underline{\tilde{B}}_i
\ee
with $\partial^i \underline{\tilde{B}}_i = 0$.  We will use a
similar decomposition for $\zeta^i$ defined in (\dr{Xi}) as well
as the usual one for tensor perturbations. Thus from
(\dr{pert00})-(\dr{pertij}) we have
\ba
\underline{A} &=& \frac{1}{n^2} \left[ \frac{\epsilon}{2} (f'- h'
\dot{R}^2) + (  \dot{\zeta}^0 f -  \dot{\epsilon}h \dot{R} )
\right],
\nn
\\
\underline{B} &=& \frac{1}{an} \left[ \zeta^0 f - \dot{\zeta} a^2 -
  \epsilon h \dot{R}\right],
\nn
\\
\underline{\tilde{B}}_i & = & - \frac{a}{n} \dot{\tilde{\zeta}}_i,
\nn
\\
\underline{C} &=& \frac{\epsilon}{2} \left( \frac{g'}{g} \right),
\nn
\\
\underline{E} &=& \zeta,
\nn
\\
\underline{\tilde{E}}_i &=& \tilde{\zeta}_i,
\nn
\\
\underline{\tilde{\tilde{E}}}_{ij} & = & 0
\nn
\ea
where we have used standard notation defined e.g.~in
\cite{Riazuelo:2002mi}. By considering coordinate transformations
on the brane and doing standard 4-dimensional perturbation theory
one can define the usual two Bardeen potentials, as well as the
brane vector and tensor metric perturbations.  For the first
Bardeen potential we find, after some algebra,
\ba
\underline{\Phi} & = & - \underline{C} + \frac{\dot{a}}{n} \left(
\underline{B} + \frac{a}{n} \underline{\dot{E}} \right)
\nn
\\
& = & \left(\frac{\dot{a}}{a} \right) \frac{f}{n^2}
\frac{1}{\dot{R}} \left[  \zeta^0 \dot{R}- \epsilon \right].
\dle{B1}
\ea
Notice that all terms containing $\zeta^i$ in $\underline{B}$ and
$\underline{E}$ have cancelled as expected since they are not
physical degrees of freedom. Similarly
\ba
\underline{\Psi} &=& \underline{A} - \frac{1}{n} \partial_t
\left(a \underline{B} + \frac{a^2}{n} \underline{\dot{E}} \right)
\nn
\\
& = & \frac{1}{n^2} \frac{1}{\dot{R}} \left[  \zeta^0 \dot{R}-
\epsilon \right] \left[ f'\dot{R} - f \left( \frac{\dot{n}}{n}
\right) \right].
\ea
The important point to notice in this second case is not only the
absence of $\zeta^i$, but that all {\it derivatives} of the
perturbations $\zeta^0$ and $\epsilon$ (which appear in
$\underline{A}$) have also cancelled.  Hence we will find that the
Bardeen potentials are proportional to $\phi$ only and not to any
of its derivatives.  Finally, the gauge invariant vector and
tensor perturbations are identically zero.

We now set
\be
\epsilon = n^4 \phi, \qquad \zeta^0 = n^0 \phi
\ee
(where $n^{\nu}$ is the normal to the brane) in order to make
contact with the covariant formalism of section
\dr{sec:perturbed}.  Then the combination which appears in both
$\underline{\Psi}$ and $\underline{\Phi}$ is
\be
    \zeta^0 \dot{R}- \epsilon = - \left( \frac{n^2(t)}{f} n^4 \right)
\phi
\ee
where $n^4$ is the 4th component of the normal to the unperturbed
brane. Thus
\be
\underline{\Phi} = - \left( \frac{\dot{a}}{a} \right)
\frac{n^4}{\dot{R}} \phi \;, \qquad \underline{\Psi} = \left(
\frac{f'}{f}\dot{R}-\frac{\dot{n}}{n} \right) \frac{n^4}{\dot{R}}
\phi
\ee
which, on going to AdS$_5$-S and using the expression for
$\dot{R}^2$ in (\dr{eq:rdotsquared}) yields
\ba
\underline{\Phi} &=& - \frac{(E-C(a))}{a^4} \left(
\frac{\phi}{\ell} \right)  =  - \left( \frac{\tE}{a^4} + q \right)
\left( \frac{\phi}{\ell} \right)  \\
\underline{\Psi} &=& 3 \underline{\Phi} + 4 q \left(
\frac{\phi}{\ell} \right).
\ea
Even though there are no anisotropic stresses, the Bardeen
potentials here  are not equal. We suppose that this is due to the
absence of self-gravity.  We see that for superhorizon modes on an
expanding brane (for which, from section \dr{sec:perturbed},
$\phi_k \propto a^4$), we also have $\underline{\Phi}_k \propto
a^4$. Similarly, $\underline{\Phi}_k$ also grows rapidly for a
brane falling into the black-hole horizon.

To obtain a true (i.e.~gauge invariant) measure of the `deviation'
from FRW, it is useful to look at the ratio of the components of
the perturbed Weyl tensor and the background Riemann tensor which
in the FRW case is roughly given by $(k\eta)^2
|\underline{\Phi}_k+\underline{\Psi}_k|$, see \dc{Durrer:1993db}.
For $\underline{\Phi}_k \propto a^4$ this ratio grows, because $a
\sim \eta^{1/3}$ when ${\cal{H}}^2 \sim a^{-6}$.


\section{Conclusions }
\label{sec:conclusions}

In this paper we have studied the evolution of perturbations on a
moving D3-brane coupled to a bulk 4-form field, focusing mainly on
a AdS$_5$-Schwarzschild bulk. For an observer on the unperturbed
brane, this motion leads to FRW expansion/contraction with scale
factor $a \propto r$. We assumed that there is no matter on the
brane and ignored the backreaction of the brane onto the bulk.
Instead, we aimed to investigate the growth of perturbations due
only to motion, and also to study the stability of moving
D3-branes.  For such a probe brane, the only possible
perturbations are those of the brane embedding. The fluctuations
about the straight brane world sheet are described by a scalar
field $\phi$ which is the proper amplitude of a `wiggle' seen by
an observer comoving with the unperturbed brane. Following the
work of \dc{Garriga:1991ts,Guven:1993ew,Buonanno:1995bf} we
derived an equation of motion for $\phi$,  and investigated
whether small fluctuations are stretched away by the expansion, or
on the other hand, whether they grow on a contracting brane.
The equation for $\phi$ is characterized by an effective mass
squared and we noted that if this mass was positive, the system is
not necessarily stable:  indeed in section \dr{sec:perturbed} we
discussed a regime in which the effective mass squared is positive
in brane time, but negative in conformal time, and therefore the
perturbations grow. Another important factor in the evolution of
$\phi$ is the time dependence of that mass.

In section \dr{sec:perturbed} we found that on an expanding BPS
brane with total energy $E=0$, superhorizon modes grow as $a^{4}$,
whereas subhorizon modes decay and hence are stable. For a
contracting brane, on the contrary, both super- and sub-horizon
modes grow  as $a^{-3}$ and $a^{-1}$ respectively. These
fluctuations become  large in the near extremal limit, $a_H \ll
1$. We therefore concluded that the brane becomes unstable (i.e.\
the wiggles grow) as it falls into the black-hole. We also
discussed the case $E>0$ for BPS branes and BPS anti-branes.
Non-BPS branes were found to be unstable at late times when a
positive cosmological constant dominates.

We have discussed the evolution of the fluctuations $\phi$ as
measured by a five dimensional observer moving with the
unperturbed brane. However, for an observer at rest in the bulk,
the magnitude of the perturbation is given by a Lorentz
contraction factor times the proper perturbation $\phi$. (For a
flat bulk spacetime this was pointed out in
\cite{Garriga:1991ts}.) Hence, if perturbations grow for the
`comoving' observer, they do not necessarily grow for an observer
at rest in the bulk.

Finally, the fluctuations around the unperturbed world sheet
generate perturbations in the FRW universe. In section
\dr{sec:bardeen} we discussed these perturbations from the point
if view of a 4D observer now living on the perturbed brane. We
calculated the Bardeen potentials $\underline{\Phi}$ and $
\underline{\Psi}$ which were both found to be proportional to
$\phi$. Furthermore, we saw that the  ratio `Weyl to Riemann'
which, expressed in terms of $\underline{\Phi}$ and $
\underline{\Psi}$, gives a gauge invariant measure for the
`deviation' from FRW, also grows.

A limitation of this work is that the back-reaction of the brane
onto the bulk was neglected. One may wonder whether inclusion of
back-reaction could stabilize $\phi$. To answer that question,
recall that the set up we have analyzed here corresponds, in the
junction condition approach, to one in which $Z_2$-symmetry across
the brane is broken. Then the brane is at the interface of two
AdS$_5$-S space times, and its total energy is related to the
difference of the respective black hole masses: $\tE \propto M_+ -
M_-$. Perturbation theory in such a non-$Z_2$ symmetric
self-interacting case has been set up in \dc{Riazuelo:2002mi},
though it is technically quite complicated. However, in the future
we hope to try to use that formalism to include the back-reaction
of the brane onto the bulk.

It would be interesting to extend this analysis to branes with $n$
codimensions: in this case one has to consider $n$ scalar fields
-- one for each normal to the brane. Formalisms to treat this
problem have been developed in \cite{Battye:1998zk,Guven:1993ex}.
In that case the equations of motion for the scalar fields are
coupled, and it  becomes a complicated task to diagonalize the
system.

Finally, it would also be interesting to consider non-zero
$F_{ab}$, and hence the effect of perturbations in the radiation
on the brane.


\section*{Acknowledgements}

We thank Ph.~Brax, E.~Dudas, S.~Foffa, M.~Maggiore, J.~Mourad,
M.~Parry, A.~Riazuelo, K.~Stelle and R.~Trotta for numerous useful
discussions and encouragement. We especially thank R.~Durrer for
her comments on the manuscript. T.B.~thanks LPT Orsay for
hospitality.

\providecommand{\href}[2]{#2}\begingroup\raggedright\endgroup

\end{document}